\documentclass[twocolumn]{article}
\usepackage{fullpage}
\usepackage{graphicx}
\usepackage{pgf}
\usepackage{booktabs}
\usepackage{url}
\usepackage{color}
\usepackage[numbers,sort&compress]{natbib}
\providecommand{\doi}[1]{doi:\ \url{https://doi.org/#1}}

\usepackage[most]{tcolorbox}
\newtcolorbox[auto counter, list inside=boxes]{sidebox}[2][]{%
  float*=tbp, width=\textwidth, colback=black!3, colframe=black!60,
  fonttitle=\bfseries, title={Box~\thetcbcounter: #2}, #1}

\title{Two blind spots in the demographic inference\\of human origins from genomic data}
\author{Ryan Gutenkunst\\
\small Department of Molecular and Cellular Biology, University of Arizona, Tucson, AZ 85721, USA}
\date{}

\begin{document}
\twocolumn[
\begin{@twocolumnfalse}
\maketitle

\begin{abstract}
\noindent
Ancient DNA and new inference methods have transformed the study of human origins, but consensus has not followed.
Evidence increasingly indicates that hominin populations were pervasively structured and admixed, so complexity rather than simplicity is the appropriate prior.
Here I highlight two blind spots that impede resolving that complexity.
First, every inference passes through summaries of the data, and those summaries bound what can be recovered.
Second, the space of candidate models is vast, yet competing model classes are rarely fit to common data, so a reported best model carries little evidence about untested model classes.
This second blind spot reflects practice rather than data.
It can be narrowed by testing competing models against withheld summaries and by reporting the models that were tried and rejected rather than only the winner.
\end{abstract}


\bigskip
\end{@twocolumnfalse}
]

\section*{Introduction}

The study of human origins draws on paleoanthropology and archaeology and increasingly on inferences of demographic history from genomic data.
Such inference aims to reconstruct past population sizes, split times, and gene flow~\cite{marchi_demographic_2021}.
Ancient DNA has transformed the field, and new methods extract more information from each genome than was possible a decade ago.
Yet consensus has not followed; the catalogue of published demographic models has grown faster than our means of choosing among them.

\begin{figure}
\centering
\includegraphics[width=\columnwidth]{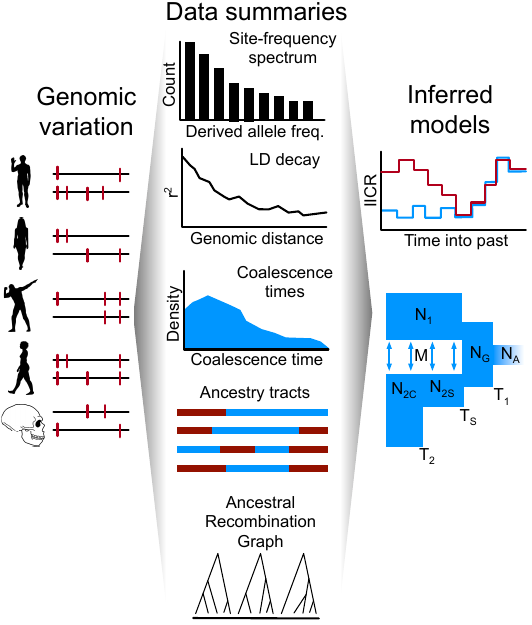}
\caption{\textbf{Every inference passes through one or more summaries.}
Genomic variation is compressed into data summaries before any demographic history is estimated.
Commonly used summaries include the site frequency spectrum, the decay of linkage disequilibrium with distance, distributions of inferred coalescence times, inferred ancestry tracts, and inferred ancestral recombination graphs.
Functional inferences estimate trajectories of instantaneous inverse coalescence rates (IICRs) through time, within and between populations.
Parametric inferences estimate parameters of an explicit model, including population sizes, divergence times, and rates of gene flow.
Either inference type can be estimated from any summary, but the summaries bound what can be recovered.
}
\label{fig:pipeline}
\end{figure}

Broadly, demographic inferences take two forms (Fig.~\ref{fig:pipeline}).
Functional approaches estimate trajectories of coalescence rates through time, within populations and between them, which can be interpreted in terms of effective population sizes, divergence, and gene flow (e.g., \cite{speidel_inferring_2021,terhorst_accelerated_2025}).
Parametric approaches instead estimate the parameters of an explicit demographic model (e.g., \cite{cousins_structured_2025}).
Either can be applied to many summaries of the data, including ancestral recombination graphs~\cite{nielsen_inference_2025} and features learned by deep neural networks~\cite{huang_harnessing_2024}.
Regardless of form, no demographic model completely captures history.
Yet key model features, such as divergence times, are interpreted to make conclusions about the past (Box~\ref{box:reading}).

\begin{sidebox}[label={box:reading}]{Reading a demographic history model}
\begin{enumerate}
\item \textbf{Check that the quantity estimated is the quantity interpreted.}
For example, functional methods estimate coalescence rates rather than census sizes (Fig.~\ref{fig:pipeline}).
\item \textbf{Determine which data summaries were used and what they cannot resolve.}
Ancient events behind a bottleneck are almost invisible to the site frequency spectrum, and small changes in data processing can dramatically alter inferences.
Every other summary has limits too, even where those limits are uncharacterized.
\item \textbf{Establish the candidate model set.}
Model selection is only meaningful within that set, and the space of plausible models is vast (Fig.~\ref{fig:modelspace}).
\item \textbf{Look for a check against something the models were not fit to.}
Good fit to the summary that was optimized is weak evidence.
Predicting a withheld summary is much stronger.
\item \textbf{Check the calibration before comparing across studies.}
Note what mutation rate, recombination rate, and generation time were used to convert from modeling units to individuals and years.
Cross-study size and time comparisons are otherwise unsafe, although unitless fractions are not affected \cite{scally_revising_2012}.
\end{enumerate}
\end{sidebox}

Tens of thousands of ancient DNA samples are available, but with important limitations in coverage and distribution.
Most approaches for demographic history inference yield times scaled in terms of mutation rate and generation time~\cite{scally_revising_2012,collmacia_different_2021a}; ancient samples provide temporal anchors to calibrate time scales, although dating can be challenging~\cite{prufer_genome_2021,sumer_earliest_2025} and extrapolation remains necessary~\cite{wohns_unified_2022,speidel_inferring_2021}.
High-coverage Denisovan~\cite{meyer_highcoverage_2012} and Neanderthal~\cite{prufer_complete_2014} genomes were transformative, but over a dozen years later, the original Denisovan genome remains the only published sample, and only five high-coverage Neanderthal genomes have been published~\cite{prufer_complete_2014,prufer_highcoverage_2017,mafessoni_highcoverage_2020,massilani_highcoverage_2026,bossomsmesa_genetic_2026}.
Ancient DNA data is available from many ancient modern human samples~\cite{mallick_allen_2024,akbari_ancient_2026}, but almost all are low coverage, preventing their use for most demographic history inference methods (though a handful of Pleistocene individuals are sequenced to high coverage~\cite{fu_genome_2014,sumer_earliest_2025}).
Moreover, ancient human samples are strongly biased toward recent times and outside of Africa~\cite{mallick_allen_2024}, leaving the critical time and geographic window for human origins largely unsampled.

This review highlights emerging consensus on the demographic complexity of human origins and two blind spots that impede resolving that complexity.
The first blind spot is that modeling approaches have fundamental limitations in what they can recover from the data, and overinterpreting model results can lead to false conclusions.
The second blind spot is that the space of candidate models is vast, and accepted procedures for navigating that space and reporting the journey toward a final model are lacking.
Other processes, including natural selection and heterogeneous mutation and recombination rates, can also bias demographic history inference \cite{marchi_demographic_2021,johri_recommendations_2022}, but here I focus on blind spots that arise even under neutrality and uniformity.

\section*{Emerging consensus}

Admixture and gene flow were common in hominin history.
The initial Neanderthal and Denisovan genomes revealed interbreeding with modern humans~\cite{green_draft_2010,meyer_highcoverage_2012}.
Later work settled on 1--2\% Neanderthal ancestry in non-Africans, and dated ancient samples have tightly constrained the timing of that admixture~\cite{fu_genome_2014,iasi_neanderthal_2024,sumer_earliest_2025}.
Denisovan ancestry reaches 3--6\% in Papuans and island Southeast Asians and arrived in multiple pulses~\cite{browning_analysis_2018,jacobs_multiple_2019}, but appropriate ancient samples are lacking and its timing remains uncertain.
Gene flow also went in other directions; Neanderthals carry roughly 3\% modern human ancestry from two pulses~\cite{hubisz_mapping_2020,li_recurrent_2024}, and Denisovan ancestry appears in some but not all sequenced Neanderthals~\cite{massilani_highcoverage_2026}.
Most directly, a first-generation Neanderthal--Denisovan hybrid has been sequenced~\cite{slon_genome_2018}, an observation which ancestral structure or incomplete lineage sorting cannot produce.

Hominin populations were also highly structured.
Genetic differentiation between Eastern and Western Neanderthals exceeded that of the most differentiated contemporary human populations~\cite{massilani_highcoverage_2026}.
Within Africa, the ancestors of the San diverged from other populations roughly 200--300~kya with substantial subsequent gene flow~\cite{fan_wholegenome_2023,breton_ancient_2026}, and ancient southern African individuals fall outside all present-day variation~\cite{jakobsson_homo_2026}.
Multiple studies also suggest even deeper structure at least 1.5~Mya~\cite{cousins_structured_2025,ragsdale_weakly_2023a}, although they disagree on the nature of that structure.

The more deeply the demographic history of human origins is understood, the more complex it becomes.
Admixture and population structure were present and perhaps pervasive.
Disentangling them is a major challenge, because they can produce similar genetic patterns \cite{tournebize_ignoring_2025}.
A tree-like model of panmictic populations is not a minor convenience; it is a major simplification \cite{chikhi_genetic_2026}.
So downstream inferences relying on demographic models must either incorporate admixture and structure or demonstrate that their results are robust to them.

\section*{Blind spot 1: What data summaries cannot resolve}

Calculating the likelihood of recombining population genomic data under even a simple demographic model is currently intractable.
So most inference methods analyze summaries of the data, such as the site frequency spectrum (SFS), linkage disequilibrium (LD) decay, or distribution of coalescence times (Fig.~\ref{fig:pipeline}).
For example, the SFS is the distribution of derived allele counts in a sample and has been widely used for demographic inference.
Different summaries retain different information, and relying on limited summaries can lead to false inferences.

In a recent high profile publication, Hu et al.~\cite{hu_genomic_2023} used their new tool FitCoal to fit the SFS of multiple African populations to infer a severe and prolonged bottleneck in the ancestors of modern humans around 900~kya.
This claim is highly contentious.
Deng et al.~\cite{deng_previously_2025} found that FitCoal inferred a sharp bottleneck from the SFS expected under a smooth model, and Cousins and Durvasula~\cite{cousins_insufficient_2025} found that simpler models fit the data SFS substantially better.
Simulation tests by Omarjee et al.~\cite{omarjee_blockbuster_2026} found that FitCoal performed well overall but could infer spurious ancient bottlenecks.
More fundamentally, the FitCoal approach assumes panmixia, which is a strong assumption given evidence for admixture and structure in hominin history.
The FitCoal authors dispute these criticisms, arguing that subtle testing details explain Deng et al.'s results, that a specific scenario of ancient structure would not produce an inferred bottleneck, and that fossil evidence supports the bottleneck \cite{zhou_severe_2026}.

Mathematical analysis has identified fundamental limitations of some summaries for demographic inference.
The power of SFS-based inference increases only slowly with the number of variant sites \cite{terhorst_fundamental_2015}, and it is sensitive to slight variations in the data \cite{rosen_geometry_2018}.
Moreover, piecewise constant histories of sufficient complexity can fit the SFS of any panmictic history \cite{rosen_geometry_2018}.
SFS-based inference has been heavily analyzed because it is mathematically tractable; similar results likely hold for other data summaries.
For example, the distribution of coalescence times between two lineages is degenerate between population structure and size changes \cite{mazet_importance_2016}.

Incorporating other data summaries can break model degeneracies, during inference or afterward.
For example, the joint distribution of first and second coalescence times among three lineages distinguishes between island models of population structure and size changes \cite{grusea_coalescence_2019}.
Notably, a recent Bayesian inference incorporating the SFS, coalescent times, and LD decay yielded little posterior support for a dramatic ancient bottleneck in human history \cite{terhorst_accelerated_2025}.
Especially powerful is testing a model by predicting data summaries the model was not fit to.
For example, Beichman et al.~\cite{beichman_comparison_2017} compared several models of human demographic history, finding that none recapitulated LD decay.
In a recent creative example, Loya et al.~\cite{loya_genomewide_2026a} inferred deep structure in the ancestors of modern humans, which they supported using evidence of differing recombination landscapes between lineages driven by differing PRDM9 alleles.

New inference methods based on ancestral recombination graphs (ARGs) and deep neural networks promise to use more of the data.
An ARG is a complete representation of the genealogical history of a sample, including local tree topologies, branch lengths, and recombination breakpoints \cite{nielsen_inference_2025}.
Neural networks learn data features to extract, either from large vectors of summary statistics \cite{quelin_assessing_2025} or directly from haplotype matrices \cite{wang_automatic_2021,huang_harnessing_2024}.
Both, however, introduce a model-dependent step between the data and the inference.
The ARG is itself a complex object \cite{wong_general_2024a}, so demographic inference methods act on summaries of the ARG \cite{dehaas_inference_2025,fan_likelihoodbased_2025}.
Early benchmarks found systematic branch-length bias even when ARG inference methods were given the true demography, mutation rate, and recombination rate \cite{brandt_evaluation_2022}.
Demographic inference is consequently often substantially worse on inferred ARGs than on true ARGs \cite{fan_likelihoodbased_2025,dehaas_inference_2025}.
Even given a true ARG, the time-stratified statistics commonly computed from it can report population structure in time windows that predate it, because allele sharing reflects how lineages later coalesced as well as when mutations arose \cite{deng_coalescentbased_2026}.
Neural networks are limited by the breadth and realism of their training simulations.
Domain adaptation can steer networks away from features that distinguish simulated from real data \cite{mo_domainadaptive_2023}, but it cannot correct misspecification of the demographic model itself.
Interpretation methods are still developing, but early work has found that neural networks may not always use more of the data than conventional summary statistics.
For example, Xu et al.~\cite{xu_interpreting_2025} found that a random forest trained on summary statistics could distinguish real from simulated genomic regions more accurately than a neural network trained on haplotype matrices.
And Tran et al.~\cite{tran_interpreting_2025} permuted test data in ways motivated by population genetics theory and showed that the demographic inference network of Flagel et al.~\cite{flagel_unreasonable_2019} relied primarily on allele frequencies.

Every demographic inference relies on summaries of the data, which bound what can be recovered.
Mathematical analysis has identified fundamental limitations of the SFS for demographic inference, and similar limitations likely hold for other summaries.
Combining summaries can break degeneracies and improve inference, and ARGs and deep learning are promising approaches to use more of the data, within their own limitations.
However, method complexity and computational expense also limit exploration of model space.
A better model that is never tried can never supplant a worse one, and the space of candidate models is vast.

\section*{Blind spot 2: Models we never fit}

\begin{figure*}
\centering
\includegraphics[width=\textwidth]{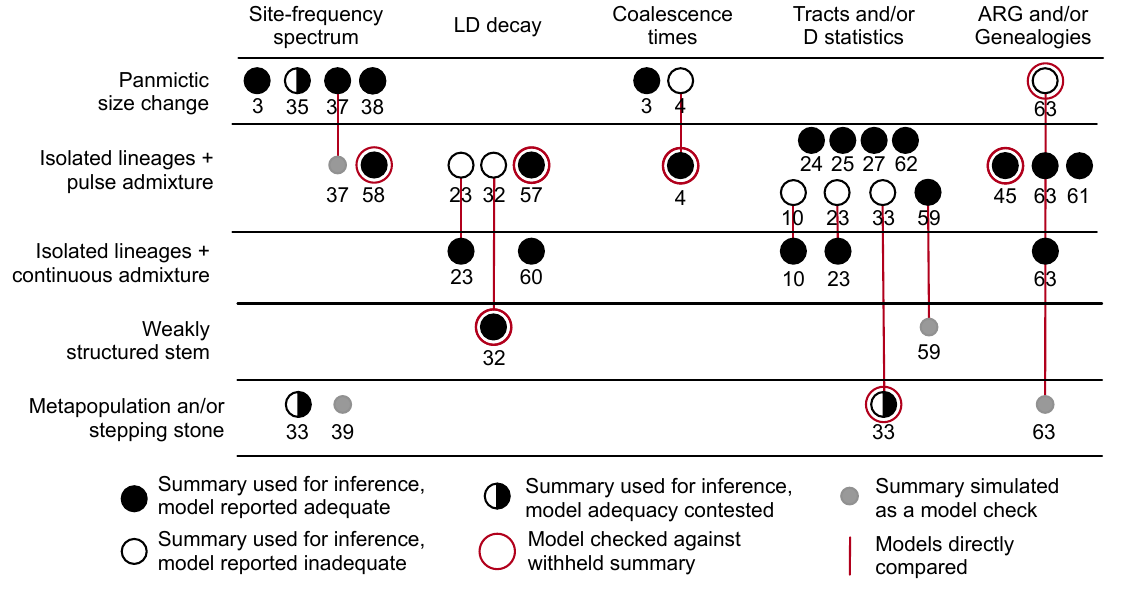}
\caption{\textbf{What model classes have been inferred from what data summaries.}
Rows are classes of demographic models and columns are data summaries.
For demographic inference papers cited in this article, circles represent what model classes were inferred, what data summaries were used, and what was concluded.
A ring indicates that the inference was checked against a withheld summary.
These papers are a select subset of the literature, chosen to illustrate progress and challenges in the demographic inference of human origins, and annotations are necessarily subjective.
}
\label{fig:modelspace}
\end{figure*}

When exploring model space, researchers must decide where to start, when to stop, and how to move through the space.
To reduce overfitting, researchers are advised to start with a simple model and add complexity as needed \cite{marchi_demographic_2021}.
Yet the emerging consensus that hominin history was pervasively structured and admixed makes a simple starting model a risky simplification rather than a neutral default.
Formal statistical criteria, such as likelihood ratio tests or the Akaike Information Criterion (AIC), can inform when to stop adding complexity.
But these criteria do not assess whether the model actually fits the data well; the practitioner must check model predictions against both summaries used to fit the model and summaries withheld from fitting \cite{johri_recommendations_2022}.
How to move through the space of models is less clear.
When growing a model, different modelers will prioritize different features based on their own priors and the limitations of the methods they are using.
The complexity of hominin history means many model features are plausible, creating many choices and many potential differing blind spots (Fig.~\ref{fig:modelspace}).

Population genetic processes are often coupled, so omitting one from a model may bias inferences of others.
Gene flow from Neanderthals into humans was quickly incorporated into models \cite{green_draft_2010}, but gene flow the other direction took longer to be established \cite{hubisz_mapping_2020}.
Incorporating that gene flow into the model reduced the estimated Neanderthal effective population size by roughly 20\% \cite{li_recurrent_2024}.
Similarly, jointly modeling two archaic introgression episodes reduced an estimated modern-human contribution to Neanderthals from 8.8\% to 5.7\%, because both events explain the same feature of the data \cite{collier_inferring_2026}.
An analysis supporting a third archaic introgression into Asia and Oceania trained on eight candidate models that all shared a fixed tree structure, never testing whether ancestral structure could produce the same signal \cite{mondal_approximate_2019}.
Similarly, unmodeled deep structure can inflate false-positive rates for introgression detection \cite{mcallister_scalable_2026}.

In hominins, confounding between signals of admixture and structure is an ongoing challenge.
A structured African metapopulation model with no Neanderthal admixture can reproduce D statistics and contemporary ancestry tract lengths attributed to Neanderthal admixture \cite{tournebize_ignoring_2025}, but not the longest Neanderthal tracts found in ancient European human samples.
Within Africa and deeper in time, ancient samples are unavailable to break the degeneracy.
Ragsdale and Gravel~\cite{ragsdale_models_2019a} were among those who inferred archaic admixture from an unsampled ghost lineage into West Africans.
Later, they expanded their set of models to include population structure within Africa, finding that model better explained the data \cite{ragsdale_weakly_2023a}.
Deeper in time, the relative roles of population structure and archaic admixture in shaping the genetic diversity of modern humans remains unresolved
\cite{cousins_structured_2025,zhang_recovering_2026a,rogers_genetic_2026,mackintosh_distribution_2026,loya_genomewide_2026a}, in part due to a lack of direct model comparisons (Fig.~\ref{fig:modelspace}).

Comparing competing model classes is difficult when models are not compatible across tools and thus data summaries, because likelihoods calculated using different data summaries are not comparable.
For example, cobraa cannot fit continuous gene flow, so Cousins et al.~\cite{cousins_structured_2025}
could not fit the weakly structured stem model~\cite{ragsdale_weakly_2023a}; conversely, the stem
model mostly omits size changes within the stems to ensure parameter identifiability and convergence.
Withheld data summaries provide a common currency for comparison, because any parametric model can be simulated to predict the same withheld summary, especially when models are represented in a common format \cite{gowerDemesStandardFormat2022}.
A hidden cost of data-rich ARG-based and deep learning methods is that they leave less of the data to withhold, impeding the sharpest model comparisons.

\begin{sidebox}[label={box:reporting}]{Building and reporting a demographic history model}
\begin{enumerate}
\item \textbf{Reserve a summary.}
Before fitting, nominate one statistic to leave out of the optimization, so an independent check is possible.
Methods that consume more of the data gain precision at the cost of leaving less to withhold.
For example, Collier et al.~\cite{collier_inferring_2026} fit two-locus statistics then checked the model by predicting coalescence-rate curves.
\item \textbf{Fit competing model classes to common data.}
Until competing classes are fit to the same data, a genuine degeneracy cannot be distinguished from an alternative that nobody has tested.
For example, Ragsdale et al.~\cite{ragsdale_weakly_2023a} did this and reattributed an archaic-admixture signal to ancestral structure, revising their earlier inference \cite{ragsdale_models_2019a}.
Mackintosh~\cite{mackintosh_distribution_2026} fit panmictic, pulse-admixture, and continuous-migration models to one set of inferred genealogies, rejecting panmictic models.
\item \textbf{Report the negative space.}
Name the plausible model classes that were not fit, and say why.
A posterior probability or AIC weight applies to the candidate set, so its interpretation depends on what that set excludes.
\item \textbf{Report a set, not a winner.}
Where several model classes are plausible, report estimates under each, and discuss difference and consensus.
For example, S\"umer et al.~\cite{sumer_earliest_2025} report estimates that are robust to continuous or pulse admixture models.
\end{enumerate}
\end{sidebox}

The second blind spot has been much less addressed than the first, because it is a limit of practice rather than of data or mathematics.
In the short term, researchers can improve the robustness of inference by directly fitting competing classes of models to common data, although this may be computationally expensive.
At little cost, they can also improve the transparency of inference by reporting not just the best-fitting model, but also the models that were tried and rejected, and by reporting a family of well-fitting models when appropriate, rather than a single winner (Box~\ref{box:reporting}).
In the long term, deeper study of the model exploration process is needed.
Automated search over model structures has shown promise \cite{noskova_gadma2_2023}, but it lacks a strong theoretical foundation.
Competitive simulation benchmarks in which the true model is complex and withheld from the modelers may also help identify best practices \cite{struck_ghist_2025}.

\section*{Conclusion}

Our understanding of human history is advancing rapidly, driven by new data and new methods.
Ancient DNA has revealed previously unknown hominin lineages, anchored the timing of demographic events, and added information unavailable from contemporary samples.
But the poor environment for preserving ancient DNA in Africa means that population genetic questions about early modern human history must be answered through modeling.
ARG-based and deep learning methods promise to increase the power of modeling by incorporating more information from the available data.

But two blind spots in the modeling process remain.
First, the data summaries used for inference have limitations.
The limitations of the SFS are known mathematically, whereas the limitations of ARG-based and deep learning methods are only beginning to be characterized.
While promising for increasing power, those methods also complicate the key process of validating a model against a withheld summary.
Second, model space is vast and difficult to explore.
When the space of models considered is too narrow, misleading inferences can arise.
This second blind spot has been much less addressed than the first and is more remediable today, because it reflects practice rather than theory.
Modelers can begin to address it by fitting competing model classes to common data, by withholding summaries for testing, and by transparently reporting the full modeling process, not just the final result.
In the long run, the field must systematically study the space of models and how best to explore it.

\section*{Acknowledgments}
Research reported in this publication was supported by the National Institute of General Medical Sciences of the National Institutes of Health under award number R35GM149235.
The content is solely the responsibility of the author and does not necessarily represent the official views of the National Institutes of Health.
I thank Aaron Ragsdale and David Castellano for helpful comments on the manuscript.

\section*{Declaration of competing interests}
The author declares that he has no known competing financial interests or personal relationships that could have appeared to influence the work reported in this paper.

\section*{Declaration of generative AI and AI-assisted technologies in the writing process}
During the preparation of this work the author used Claude (Anthropic) while drafting text and figures and for formative feedback.
After using this tool, the author reviewed and edited the content as needed and takes full responsibility for the content of the publication.

\bibliography{Papers/Gutengroup}
\end{document}